\documentclass[prd,twocolumn]{revtex4}
\usepackage{amssymb}
\usepackage{color}
\usepackage{graphicx}
\usepackage{comment}

\def\lsi {LS~I~+61$^\circ$~303}
\def\gr{$\gamma$-ray}

\begin{document}
\title{Neutrino signal from $\gamma$-ray loud binaries powered by high energy protons}
\author{A.Neronov$^{1,2}$, M.Ribordy$^3$}
\address{
$^1$ INTEGRAL Science Data Centre, 1290 Versoix, Switzerland; \\
$^2$ Geneva University Observatory, 1290 Sauverny, Switzerland;\\
$^3$ High Energy Physics Laboratory, EPFL, 1015 Lausanne, Switzerland}
\begin{abstract}
We present a hadronic model of activity for Galactic \gr-loud binaries, in which the multi-TeV neutrino flux from the source can be much higher and/or harder than the detected TeV \gr\ flux. This is related to the fact that  most neutrinos are produced in pp interactions close to the bright massive star, in a region optically thick for the TeV \gr s. Considering the specific example of \lsi, we derive upper bounds for neutrino fluxes from various proton injection spectra compatible with the observed multi-wavelength spectrum. At this upper level of neutrino emission, we demonstrate that ICECUBE will not only detect this source at $5\sigma$ C.L. after one year of operation, but, after 3 years of exposure, will also collect a sample marginally sufficient to constrain the spectral characteristics of the neutrino signal, directly related to the underlying source acceleration mechanisms.
\end{abstract}
\maketitle

\section{Introduction}
The recently discovered "\gr -loud" binaries (GRLB) form a new sub-class of Galactic binary star systems which emit GeV-TeV \gr s  \cite{ls5039,lsi,psrb, HESSJ0632,cygx1}. These are high mass X-ray binaries (HMXRB) composed of a compact object (a black hole or a neutron star) orbiting a massive star. The detection of \gr s with energies up to 10~TeV from these systems shows that certain HMXRBs host powerful particle accelerators producing electrons and/or protons with energies above 10~TeV.   

At the moment it is not clear which physical process leads to the very high energy (VHE) particle acceleration and whether the VHE particle acceleration is a generic feature of the HMXRBs or the result of specific physical conditions in a restricted HMXRB sub-class. It is possible that  particle acceleration is taking place in a large number of X-ray binaries, but the \gr s can be detected only from the objects with preferred orientations w.r.t. the line of sight (e.g. with jets pointing toward the observer) \cite{mirabel}. Alternatively, it may be that the conventional accretion-powered binaries are not capable of accelerating particles and that the \gr\ activity of a binary is powered by a different mechanism (see e.g. \cite{maraschi,tavani97}).   

The theoretical models of \gr\ activity of HMXRBs \cite{maraschi,tavani97,bosch-ramon,dubus,bednarek,khangulyan,romero,chernyakova06a,chernyakova06b} all assume \gr\ emission  from the interaction of a relativistic outflow from the compact object (jet from a black hole, or wide angle wind from a pulsar) with the wind and radiation emitted by the companion massive star. The basic properties of the outflows, such as the composition (e$^+$e$^-$ pairs or electron-nuclei plasma), anisotropy (a collimated jet or a wide angle outflow) etc. are as yet poorly constrained by the data.

In one of the three persistent (periodic) \gr-loud binaries, PSR B1259-63, the relativistic outflow is known to be produced by a young pulsar.  In principle, a similar mechanism could power the activity of other sources (except for Cyg X-1), although direct proof of the presence of the young pulsar in these systems is not possible: the radio emission from the compact object is free-free absorbed. Recent puzzling detection of a short soft \gr\ flare from one of the \gr-loud binaries, LSI +61 303 \cite{lsi_burst}, may indicate the presence of a neutron star in the system.  The outflows from LSI +61 303 and LS 5039, extending to the distances $\sim 10^{14}$~cm (far beyond the binary orbit) are revealed by the radio observations \cite{dhawan,ribo08}. The nature of the outflows (a collimated jet or a wide-angle outflow) is still debated. 

With the exception of Cyg X-1, all the known GRLBs have similar spectral energy distributions (SED), peaking in the MeV-GeV energy band. The physical mechanism producing the MeV-GeV bump in the spectra is, however, not clear. The \gr\ emission from GRLBs is supposed to come from internal and/or external shocks formed in the relativistic outflow either as a result of the development of intrinsic instabilities or through interactions with the stellar wind of the massive star.
The MeV-GeV emission can be the synchrotron emission from electrons with energies much above TeV \cite{tavani97,bosch-ramon}, produced locally at the shock, or else, be produced via inverse Compton scattering of the UV thermal emission from the massive star by electrons of the energies $E\sim 10$~MeV \cite{maraschi,chernyakova06a,chernyakova06b}.

The available multi-wavelength data do not constrain the composition of the relativistic wind from the compact object. On the one hand, the multi-TeV or 10~MeV electrons, responsible for the production of the MeV-GeV bump in the SED, could be injected into the shock region from the e$^+$e$^-$ pairs loaded wind. On the other, these electrons could be secondary particles produced in e.g. proton-proton collisions, if the relativistic wind is proton-loaded. The only direct way to test if relativistic protons are present in the \gr\ emission region would be the detection of multi-TeV neutrinos.

The possibility of detection of neutrinos from GRLBs by a km$^3$-class neutrino telescopes was considered in several references \cite{aharonian07n,aharonian06n,orellana,halzen,kappes,christiansen}. If the sources are assumed to be transparent to the TeV \gr s, the estimation of the neutrino flux is straightworward, given a known source \gr\ flux and spectrum. The assumption of source transparency was adopted e.g. in the estimates of the number of detected neutrinos as in Ref. \cite{kappes}. However, the TeV \gr\ flux from the GRLBs can be significantly attenuated by the pair production on the UV photon field of the massive star \cite{dubus,bednarek}. In addition, if the \gr s and neutrinos are produced close to the compact object, the \gr\ flux can be further suppressed by pair production on the soft photons emitted by the accretion flow \cite{aharonian06n}. The derivation of the estimate of the neutrino flux and spectral characteristics based on the observed TeV \gr\ emission is inconclusive due to the uncertain attenuation of the \gr\ flux in the TeV band (see \cite{halzen} for s specific discussion of \lsi). 

In the absence of a direct relation between the characteristics of the observed TeV \gr\ and the neutrino emission from a GRLB, the only way to constrain possible neutrino signals from the source is via the detailed modelling of the broad-band spectrum of the source within the hadronic model of activity. The idea is that the pp interactions, which result in the production of neutrinos, also result in the production of e$^+$e$^-$ pairs and the subsequent release of their energy via synchrotron, inverse Compton and Bremsstrahlung emission. The electromagnetic emission from the secondary e$^+$e$^-$ pairs is readily detectable. The total power released in the pp interactions determines the overall luminosity of the emission from the secondary pairs. The known electromagnetic luminosity and broad-band spectral characteristics of the source can be used to constrain the power released in pp interactions as well as the spectrum of the primary high energy protons. 

In the following we develop the hadronic model of activity of GRLBs and derive the constraints on the spectrum and overall luminosity of neutrino emission from the analysis of the broad-band spectral characteristics of GRLBs.  Although the following discussion is generically applicable for the GRLBs as a class, we concentrate on the particular example of the \lsi\ system, because it is the only known persistent GRLB in the Northern hemisphere, available for observations with the ICECUBE neutrino telescope \cite{icecube}. 
The existing observational data are found to be consistent with a possible very strong neutrino emission from \lsi\ (with a flux at the level of $10^{-10}$~erg/cm$^2$s), close to the best reported AMANDA upper limit, $\Phi(1.6\mbox{ TeV}<E_\nu<2.5\mbox{ PeV})\le 1.26\times 10^{-10}\mbox{ cm}^{-1}\mbox{ s}^{-1}\mbox{ TeV}^{-1}$) \cite{amanda} (a more recent reference \cite{a2-7yrs} combining more data reported a slightly degraded upper limit for \lsi\ over approximately the same range of energy, while the sensitivity had been improved by roughly a factor two; this however translates a non significant but large background fluctuation at the \lsi\ source location).
Moreover, the assumption of an almost arbitrarily hard neutrino spectrum (e.g. a powerlaw $dN_\nu/dE\sim E^{-\Gamma_\nu}$ with index $\Gamma_\nu\sim 0$) is not ruled out by the available multi-wavelength observational data. This means that, in principle, the soure could be readily detectable  with ICECUBE.

The plan of the paper is as follows: In section \ref{sec:model} we describe the general features of the hadronic model of activity of GRLBs. In particular, we stress that when the massive companion star is a Be type star, the presence of a dense equatorial decretion disk around the massive star can boost the pp interaction rate. At the same time, the \gr\ emission from the pp interactions in the disk would be strongly suppressed, because of the large density of the soft photon background in the direct vicinity of the star. In section \ref{sec:numeric} we perform the detailed numerical modeling of the broad-band spectra of GRLBs in the hadronic model and show, on the particular example of \lsi, that the model provides a suitable explanation of the typical shape of the GRLB SEDs. Satisfactory fits to the observed SEDs can be achieved under quite arbitrary assumptions about the shape of the initial high energy proton spectrum. In particular, the initial spectra as different as an $E^{-2}$ power law with a high energy cut-off and a monochromatic proton spectrum peaked at $\sim$PeV energy could both explain the observed X-ray-to-\gr\ spectrum of the source. Finally, in section \ref{sec:icecube} we work out the predictions for the detection of neutrinos from \lsi\ with ICECUBE. We show that in an optimistic scenario, when the anisotropy of the neutrino emission of the source does not result in a suppression of the neutrino flux in the direction of the Earth, \lsi\ should be detectable within a year of exposure. Nevertheless, the spectral characteristics of the emission can only be marginally delineated after three years of exposure, given the wash out of the original neutrino spectrum from the measurement of the muon spectrum.

\section{Hadronic model of \gr\ activity}
\label{sec:model}
\subsection{Origin of high energy protons}

In the hadronic model, the primary source of the system's high energy activity are high energy protons. The presence of the high energy protons in relativistic outflows from compact objects (stellar mass and supermassive black holes, neutron stars) is usually difficult to detect, because of their very low energy loss rates. For example, in the case of relativistic winds ejected by young pulsars, it is possible that most of the wind power is carried by the high energy protons or ions, but the presence of protons/ions in the wind can be established only indirectly, because most of the radiation detected from a nebula powered by the pulsar wind is emitted by electrons accelerated at the shock whose properties are determined by the parameters of the proton/ion component of the pulsar wind  \cite{gallant94}. \gr\ emission initiated by interactions of the pulsar wind protons with the ambient medium can be detected only if the density of the medium is high enough \cite{horns06,bednarek03}. 

GRLBs provide a unique opportunity to "trace" the presence of protons/ions in the relativistic outflows generated by compact objects. The dense matter and radiation environment created by the companion massive star provides abundant target material for the protons in the relativistic outflow. Interactions of high energy protons with the ambient matter and radiation fields, created by the presence of a bright massive companion star, lead to the production and subsequent decays of pions. This results in the emission of neutrinos and \gr s  from the source and to the deposition of e$^+$e$^-$ pairs throughout the proton interaction region. Radiative cooling of the secondary e$^+$e$^-$ pairs leads to the broad-band synchrotron and inverse Compton emission from the source.

If the primary high energy protons are produced close to the compact object, the maximal energies of particles accelerated in the region of the size $R_{\rm c}$ in magnetic field $B_{\rm c}$, can be estimated as 
\begin{equation}
E_p\sim \kappa eB_{\rm c}R_{\rm c}\simeq 10^{15}\kappa \left[\frac{B_{\rm c}}{10^7\mbox{ G}}\right]\left[\frac{R_{\rm c}}{10^{6}\mbox{ cm}}\right]\mbox{ eV}
\end{equation}
The parameter $\kappa\le 1$ characterizes the acceleration process efficiency. For example, in the case when the compact object is a rotation-powered neutron star, $B\sim 10^{12}$~G, while $\kappa\sim \Omega_{\rm c}^2R_{\rm c}^2/c^2\sim 4\times 10^{-6}\left[P_{\rm c}/0.1\mbox{ s}\right]^{-2}$, with $R_{\rm c}, \Omega_{\rm c}, P_{\rm c}$ being respectively the neutron star radius, angular velocity and rotation period.
Depending on the particular acceleration mechanism, the proton energy spectrum below this maximal energy can range from the conventional power law shape with spectral index close to $\Gamma_p\simeq 2$ (acceleration in non-relativistic shocks), to almost monochromatic spectra (if the acceleration proceeds in large scale electric fields in the magnetic reconnection regions of accretion flow, or in the vacuum gaps in black hole or pulsar magnetosphere).

Otherwise, protons can be accelerated via shock acceleration in an extended region of the size $R_{\rm ext}$ with magnetic field $B_{\rm ext}$ in a jet or in a shocked relativistic wind emitted by the compact object. In this case one expects a conventional $E^{-2}$ type spectrum with a cut-off at the energy at which the Larmor radius of the high energy particles becomes comparable to the size of the system
$E_p\simeq 0.3\times 10^{15} \left[B_{\rm ext}/1\mbox{ G}\right]\left[R_{\rm ext}/10^{12}\mbox{ cm}\right]\mbox{ eV}$.

\subsection{Interactions of high energy protons}

Efficient interaction of high energy protons with the radiation field produced by the bright massive star in the system (e.g. a Be star with temperature $T_*\sim 3\times 10^4$~K in the case of \lsi\ and PSR B1259-63) occurs above $E_p\ge [200\mbox{ MeV}/\epsilon_*]m_p\simeq 2\times 10^{16}\left[\epsilon_*/10\mbox{ eV}\right]$~eV, where $\epsilon_*\simeq 3kT_*$ is the typical photon energy of the stellar radiation. In contrast, high energy protons efficiently interact with the protons from the dense stellar wind already at much lower energies.

In the case of massive stars of type Be, the pp interaction rate can be highly enhanced if the high energy protons are able to penetrate the dense equatorial disk known to surround them. An obstacle for the penetration of the high energy protons accelerated e.g. close to the compact object into the disk could be the presence of magnetic fields, which would deviate the proton trajectories away from the disk. However, the Larmor radius of the highest energy protons, $R_L\simeq 4\times 10^{12}\left[E_p/10^{15}\mbox{ eV}\right]\left[B/1\mbox{ G}\right]^{-1}$~cm, could be comparable to the size of the system. Thus, if the magnetic field in the region of contact between the stellar wind and the relativistic outflow is not larger than several Gauss, the highest energy protons can freely penetrate the dense stellar wind region. Note that the same effect may also result in the production of very hard injection spectra of the high energy protons in the pp interaction region (penetration of the lower energy protons into the dense stellar wind suppressed by the magnetic fields).

The estimation of the pp interaction rate in the disk is straighforward, provided a known disk density profile. For example, in the case of the Be star in the \lsi\ system, the density profile of the equatorial disc has been measured, using the IR free-free and free-bound radiation (see e.g. Refs. \citet{waters88,mp95}). Assuming a disc half-opening angle of $\theta_0\sim 15^\circ$ and a stellar radius of $R_*=10 R_\odot$, the disk density profile can be modelled as a power law of the distance $D$
\begin{equation}
n_{\rm d}(D)\simeq n_{\rm d,0} \left(D\over R_*\right)^{-\gamma},
\label{n_disc}
\end{equation}
where $n_{\rm d,0}\sim 10^{13}$ cm$^{-3}$ and $\gamma\simeq 3$. 

Given $\sigma_{pp}\simeq 10^{-25}$~cm$^2$, the pp interaction cross-section at $E_p\sim 10^{15}$~eV, one finds the characteristic time of pp interactions in the disk 
\begin{equation}
\label{tpp}
t_{pp}(D)=\frac{1}{\sigma_{pp}n_{\rm d}}\simeq 30\left[\frac{D}{R_*}\right]^\gamma\mbox{ s}
\end{equation}

To estimate the efficiency of the pp interactions, the pp interaction time should be compared to the escape time from the system. Since the details of the escape regime are not known, one can only compare (\ref{tpp}) with possible relevant time scales. The shortest possible time scale is set up by the light-crossing time of the system,
\begin{equation}
t_{\rm lc}=\frac{D}{c}\simeq 20\left[\frac{D}{R_*}\right]\mbox{ s}
\label{t_lc}
\end{equation}
This time scale is relevant e.g. if the Larmor radius of the high energy protons is comparable to the size of the disk, so that these protons cannot be trapped in the disk interior.
If the high energy protons escape on this shortest time scale, the pp interactions can only be efficient in the direct vicinity of the Be star, at $D\sim 10^{12}$~cm. 

The escape of the high energy particles can be slowed down in the presence of a sufficiently strong  tangled magnetic field $B$. In this case, the characteristic escape time is given by the diffusion time which is, in the Bohm diffusion approximation 
\begin{equation}
t_{\rm diff}=\frac{3e B D^2}{2E_p c}\simeq 74\left[\frac{D}{R_*}\right]^2
 \left[\frac{B}{1\,\mbox{G}}\right]\left[\frac{10^{14}\,\mbox{eV}}{E_p}\right] \mbox{s},
\label{t_diff}
\end{equation}
Even if the high energy particles are trapped by the tangled magnetic fields in the stellar wind, they do not stay in the innermost dense stellar wind region for a long time: the high energy particles escape together with the stellar wind of velocity $v_{\rm w}$, so that the escape time is limited from above by
\begin{equation} 
  t_{\rm wind}\sim D/v_{\rm w}\sim 7\times 10^3\left[\frac{D}{R_*}\right]\left[\frac{v_{\rm w}}{10^8\,\mbox{cm s}^{-1}}\right]^{-1}\,\mbox{s.} 
\label{tesc}
\end{equation} 
In the equatorial disk of a Be star, the initial wind velocity close to the surface of the star is believed to be slow with $v_{\rm w}\sim 1-20$~km/s close to the surface of the star  \citep{waters88,mp95,porter98}.  The polar component of the wind is faster and less dense than the equatorial one. Both equatorial and polar winds accelerate with distance, reaching asymptotically the velocity $v_{\infty}\simeq (1$--$2)\times 10^3$ km s$^{-1}$. Comparing (\ref{tesc}) with (\ref{tpp}) one finds that the pp interactions in the Be star disk can efficiently transfer the power contained in the high energy protons into neutrinos and \gr s. 

\subsection{Broad-band emission from pp interactions}

pp interactions result in the production of pions, which subsequently decay onto \gr s, neutrinos and electrons/positrons. The neutrinos freely escape from the source. The electrons/positrons suffer synchrotron and inverse Compton (IC) energy loss and, in this way release their energy through \gr\ emission. 

\begin{figure} 
\centerline{\includegraphics[width=0.9\linewidth]{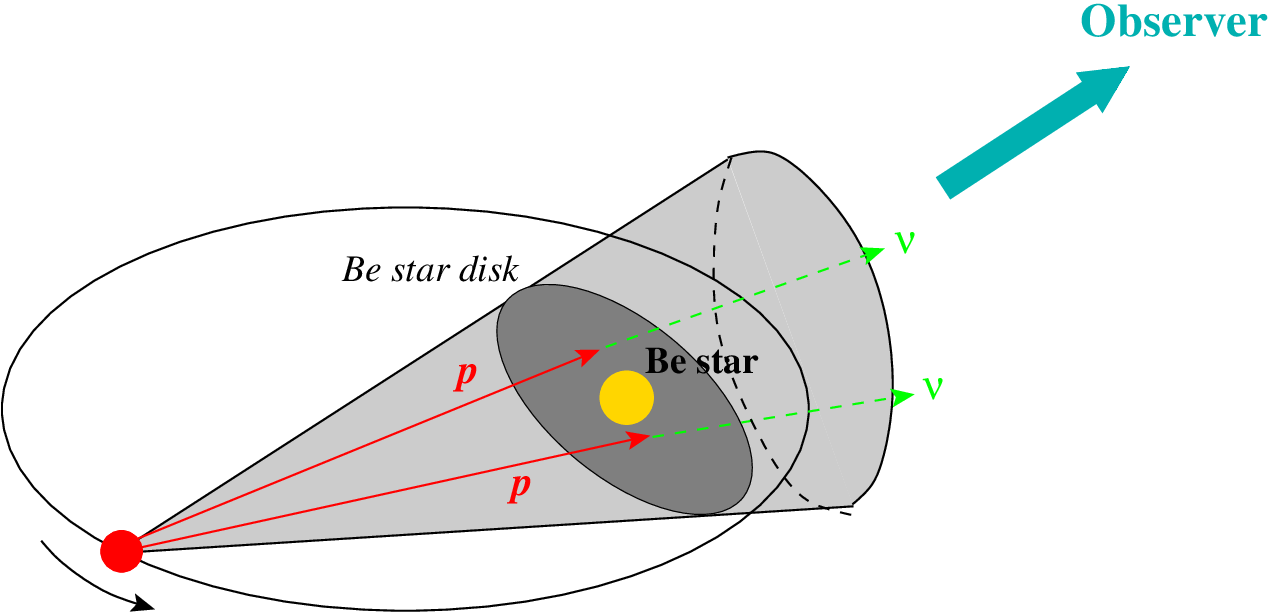}}
\caption{Mechanism of production of neutrinos in interactions of high energy protons ejected by the compact object with the dense equatorial disk of Be star. }
\label{fig:ellipse} 
\end{figure}

If the Larmor radius of the highest energy protons is comparable to the size of the disk, most of the pions are produced by the protons propagating toward the companion star.  In this case the neutrino and \gr\ emission from the pion decays is expected to be anisotropic, with most of the neutrino flux emitted toward the massive star, as it is shown in Fig. \ref{fig:ellipse}. Conversely, the synchrotron and IC emission from the e$^+$e$^-$ pairs is, most probably, isotropized at energies at which the radiative cooling time of electrons becomes longer than the period of gyration in the magnetic field. Differing anisotropy patterns for neutrino emission and for broad-band emission from the secondary e$^+$e$^-$ pairs should, in principle, lead to significant differences in the expected orbital lightcurves of neutrino and electromagnetic emission from the source. 

The IC cooling time of e$^+$e$^-$ pairs is shortest at the boundary between the
Thomson and Klein-Nishina limits for Compton scattering, $3kT_\star E
\sim (m_{\rm e}c^2)^2$, when $E\simeq 30$ GeV:
\begin{eqnarray}
t_{\rm IC}&=&\frac{3\pi m_{\rm e}^2 c^4 D^2}{\sigma_{\rm T} L_\star E}\nonumber\\ &\simeq&
2.5\left[\frac{10^{38}\,\mbox{erg\,s}^{-1} }{L_\star}\right]
\left[\frac{D}{R_*}\right]^2
\left[\frac{10^{10}\,\mbox{eV}}{E}\right]\,\mbox{s},
\label{t_IC}
\end{eqnarray} 
At the same time, the period of rotation around the Larmor circle is $t_L=2\pi R_L/c\simeq 1.3\left[E/10^{10}\mbox{ eV}\right]\left[B/1\mbox{ G}\right]^{-1}$~ms$\ll t_{\rm IC}$.

The IC scattering on higher energy electrons proceeds in the Klein-Nishina regime. In
this regime, $\epsilon\simeq E$, and the electron energy loss time
grows with energy just slightly slower than the Larmor radius \citep{bg70}
\begin{eqnarray}
\label{tkn}
t_{\rm KN} &\simeq& {2 E D^2 h^3\over \sigma_{\rm T}\pi^3(m_{\rm e} c k T_\star R_\star)^2 }\ln^{-1}{0.552 E k T_\star\over m_{\rm e}^2 c^4}\nonumber\\
&\simeq& 5\times 10^2
\left[\frac{E}{10^{14}\,\mbox{eV}}\right]
\left[\frac{D}{R_*}\right]^2\mbox{s,}
\end{eqnarray}
where the density of the blackbody photons has been diluted by the $(D/R_\star)^2$ factor. This means that if the dominant mechanism of radiative cooling is the IC energy loss, the emission from the secondary e$^+$e$^-$ pairs is isotropic.

The synchrotron cooling time,
\begin{equation}
\label{tsynch}
t_{\rm S}=\frac{6\pi m_{\rm e}^2 c^3}{\sigma_{\rm T} 
B^2 E}\simeq 4  \left[\frac{1\,\mbox{G}}{B}\right]^2
\left[\frac{10^{14}\,\mbox{eV}}{E}\right]\,\mbox{s}
\end{equation}
can be much shorter than the period of rotation around the Larmor circle only at the highest energies
\begin{equation}
E>\frac{\sqrt{3}m_ec^2}{(\sigma_T B)^{1/2}}\simeq 6\times 10^{13}\left[\frac{B}{1\mbox{ G}}\right]^{-1/2}\mbox{ eV}
\end{equation}
Taking into account that the synchrotron emission from electrons of energy $E$ is emitted in the energy band
\begin{equation} 
\label{esynch}
\epsilon_{\rm S}=\frac{eBE^2}{2\pi m_{\rm e}^3 c^5}\simeq
3\left[\frac{B}{1\,\mbox{G}}\right]\left[ \frac{E}{10^{13}\,\mbox{eV}}\right]^{2}\mbox{ MeV}, 
\end{equation} 
this emission is expected to be isotropic below several MeV.

Comparing the synchrotron and IC cooling times at energies $E\ll 30$~GeV and $E\gg 30$~GeV, one finds that the lower energy electrons predominantly dissipate their energy via inverse Compton in the energy band
\begin{equation}
\label{eic}
\epsilon_{\rm IC}\simeq 3 k T_*\left(\frac{E}{m_{\rm e} c^2}\right)^2\simeq
3\left[\frac{T_\star}{3\times 10^4\,\mbox{K}}\right]
\left[\frac{E}{10^{10}\,\mbox{eV}}\right]^{2}\mbox{GeV},
\end{equation}
while synchrotron radiation losses dominate for the highest energy electrons.

A significant contribution to the cooling of electrons and positrons in the densest part of the pp interaction region may arise from Bremsstrahlung emission, with cooling time $t_{\rm Brems}\simeq 10^2\left[n_d/10^{13}\mbox{ cm}\right]^{-1}$~s, which compares to the synchrotron / IC cooling time of $\sim$TeV electrons.

\subsection{Absorption of 10~GeV-TeV \gr s}

The flux of \gr s from the pp interaction region is absorbed due to the pair production on the UV photon background in the vicinity of the Be star.  Maximal optical depth with respect to the pair production is achieved at energies $E_\gamma\simeq 4m_e^2/\epsilon_*\simeq 0.2\left[T_*/3\times 10^4\mbox{ K}\right]^{-1}$~TeV, where the pair production cross section reaches its maximum $\sigma_{\gamma\gamma}\simeq 1.5\times 10^{-25}$~cm$^2$.  The density of photons close to the surface of the Be star is $n_{\rm ph}\simeq 5\times 10^{14}\left[T_*/3\times 10^4\right]^3\left[D/R_*\right]^{-2}$~cm$^{-3}$, thus an estimate for the optical depth for \gr s of this energy is given by 
\begin{equation}
\tau_{\gamma\gamma}\simeq \sigma_{\gamma\gamma}n_{\rm ph}D\simeq 40\left[\frac{T_*}{3\times 10^4\mbox{ K}}\right]^3\left[\frac{D}{R_*}\right]^{-1},
\end{equation}
assuming the radius of the star $R_*\simeq 5\times 10^{11}$~cm.
At higher \gr\ energies, $E_\gamma T_*\gg (m_{\rm e}c^2)^2$, the pair production cross-section and the optical depth decrease as $E^{-1}\mbox{ln}E$. The attenuation of the \gr\ flux due to the pair production becomes small only at energies above $\sim 10$~TeV. Below this energy, the spectrum of the \gr\ emission can be significantly different from that of the neutrino emission, so that no solid prediction for the neutrino flux can be derived based on the observed \gr\ flux and spectrum in the TeV band. 

The power of the absorbed \gr s is re-distributed to the secondary e$^+$e$^-$ pairs, which subsequently loose their energy through synchrotron and inverse Compton emission. Depending on the magnetic field strength in the pair production region, the bulk of the electromagnetic emission from the secondary pairs with energies ranging approximately between 10~GeV and 10~TeV can be either re-emitted back in the GeV-TeV energy band, if the inverse Compton loss dominates, or in the X-ray band, in case the synchrotron losses dominate (see Eqs. (\ref{esynch}), (\ref{eic})). 

\section{Numerical modelling of the broad-band spectrum}
\label{sec:numeric}

In order to demonstrate the possibility of dramatically different spectra of neutrino and \gr\ emission from the system in the TeV energy band, we have modeled the emission from pp interactions, assuming different proton injection spectra.
Our numerical code follows the evolution of the secondary particle spectra produced by the interaction of the high energy protons with the stellar wind protons in the course of their escape from the system. The equatorial wind of the massive star is supposed to have the radial density profile described by Eq. (\ref{n_disc}). We assume that the high energy protons are initially injected at a distance $D_{\rm min}\simeq 1.2R_*$ and then escape together with the secondary particles produced in pp interactions toward larger distances. We explore the different options for the escape regime, as described in Section \ref{sec:model}.

One possibility occurs when the high energy protons are traversing the pp interaction region without being trapped by the magnetic fields, the escape velocity is comparable to the speed of light. A similar "fast escape" situation is present when the magnetic field in the contact region between the stellar wind and the relativistic outflow is ordered (the situation present e.g. in the scenario of interaction of relativistic pulsar wind with the Be stellar wind). To model this "fast escape" regime, we assume that the primary high energy protons and the secondary e$^+$e$^-$ pairs move toward larger distances with speed $v_{\rm esc}\sim c$.

Another possibility is that the high energy protons, which penetrate into the stellar wind, and/or the secondary e$^+$e$^-$ pairs, deposited throughout the stellar wind in result of pp interactions, 
are trapped by the tangled magnetic field in the stellar wind. In this case they would escape with velocity $v_{\rm esc}\sim v_{\rm w}(D)$ approximately equal to the stellar wind velocity. The slow down of escape mostly affects the low energy parts of the electron spectra, at energies at which the radiative cooling time is comparable or larger than the escape time from the system.

We calculate the production spectra of neutrinos, \gr s and e$^+$e$^-$ pairs using the approximations given in the Ref. \cite{kelner}. The synchrotron, IC and Bremsstrahlung emission from the secondary e$^+$e$^-$ pairs, as well as the evolution of the spectra of the pairs due to the radiative cooling effects is modeled in a standard way \cite{bg70}, via a solution of the kinetic equations with the derivative over the distance $v_{\rm esc}\partial/\partial D$ substituted for the time derivative $\partial/\partial t$. 

To model the synchrotron emission, we assume a certain radial profile of the magnetic field, $B=B_0\left(D/R_*\right)^{-\alpha_B}$ with $\alpha_B=1$. The parameters $B_0$ and $\alpha_B$, used for numerical calculations shown in Figs. \ref{fig:without_pion}, \ref{fig:gamma0} and \ref{fig:gamma1} is $B_0=5$~G, while for the calculation of Fig. \ref{fig:gamma2},  $B_0=0.5$~G. Taking into account that the trajectories of the secondary e$^+$e$^-$ pairs are isotropized by the magnetic fields and that we are interested in the orbit-averaged spectrum of the source, we take the angle-averaged cross-section of IC scattering for the calculation of the IC emission.

\subsection{Broad-band spectrum of a \gr -loud binary in hadronic interaction model}

\begin{figure} 
\centerline{\includegraphics[width=0.9\linewidth]{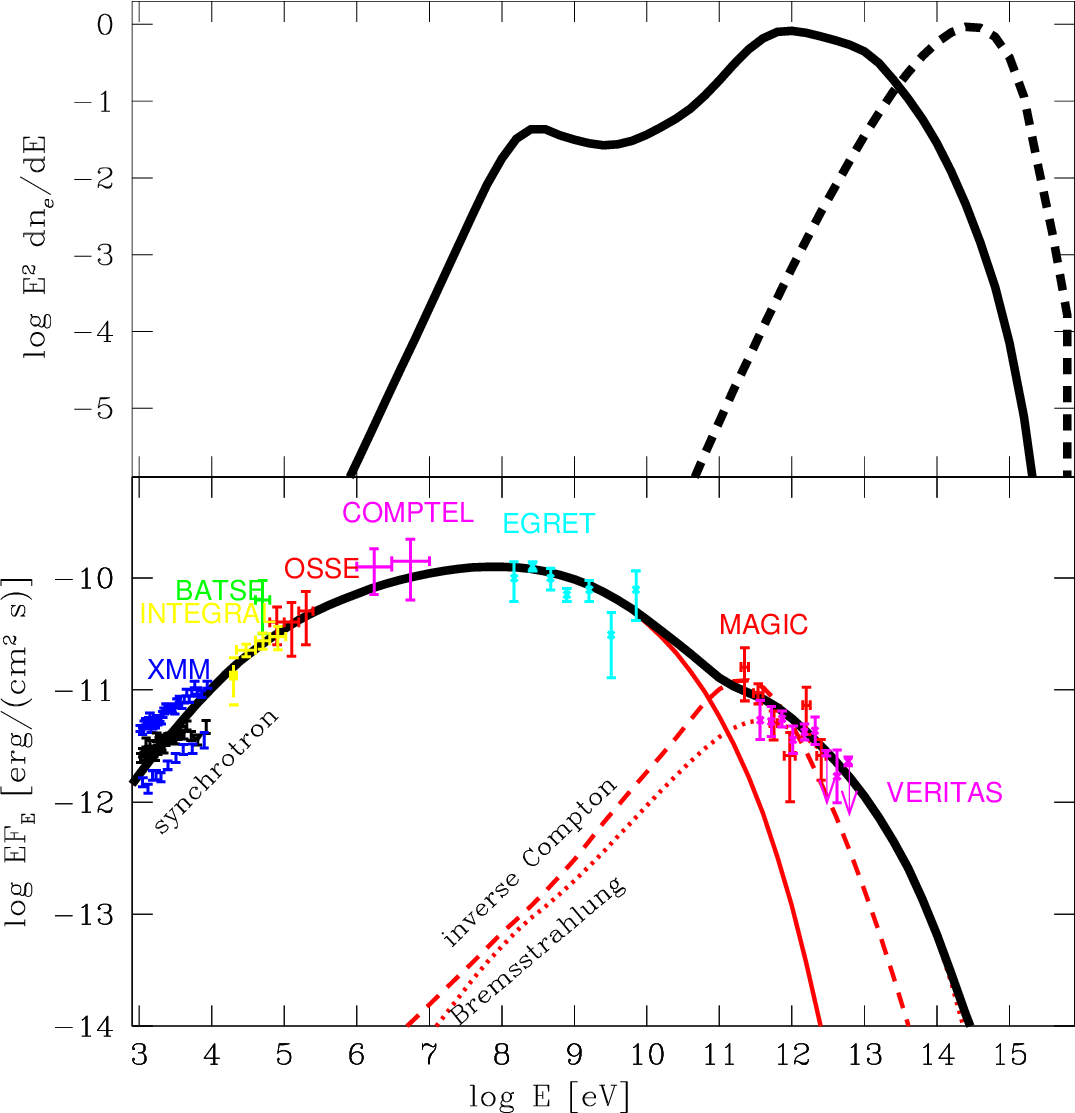}}
\caption{Broad-band spectrum of emission from secondary e$^+$e$^-$ pairs produced in pp interactions close to the surface of the Be star, calculated assuming a nearly monochromatic proton injection spectrum with $\Gamma_p=0$,  $E_{p,\rm cut}=10^{15}$~eV. The upper panel shows the injection spectrum of e$^+$e$^-$ pairs (dashed line) and the spectrum formed as a result of cooling via synchrotron, IC and Bremsstrahlung emission as well as Coulomb energy loss. In the lower panel, thin red solid, dashed and dotted lines show respectively the synchrotron, the IC and the Bremsstrahlung emission from the  pairs. The black thick solid line shows the overall broad-band model spectrum.}
\label{fig:without_pion} 
\end{figure}

The broad-band spectra of the three known persistent or periodic \gr\ loud binaries, LSI +61 303, LS 5039 and PSR B1259-63 have similar shape (which does not resemble the shape of typical accretion-powered X-ray binaries \cite{chernyakova06}). The \gr -loud binaries spectral energy distribution are peaked in the MeV-GeV energy band. The observed X-ray and TeV emission apparently form the low and high energy tails of the MeV-GeV "bump" in the spectrum. Fig. \ref{fig:without_pion} shows an example of LSI +61 303 SED.

Taking into account the possible anisotropy of the neutrino and neutral pion decay emission, we first show in Fig. \ref{fig:without_pion} the broad-band isotropic emission spectrum from the secondary e$^+$e$^-$ pairs. The primary proton injection spectrum is assumed to be hard, with $\Gamma_p\simeq 0$, so that most of the protons have the energy close to the cut-off energy assumed to be $E_{\rm cut}=10^{15}$~eV. As discussed above, such an almost monochromatic spectrum of protons injected into the stellar wind can be produced when the high energy protons originate from a "cold" relativistic wind with bulk Lorentz factor $\sim 10^6$ or when only the highest energy protons have large enough Larmor radii to be able to penetrate deep into the stellar wind. 

Within the hadronic model of activity, the SED MeV-GeV bump can be ascribed to the synchrotron emission from the e$^+$e$^-$ pairs produced in the pp and $\gamma\gamma$ interactions. One should note that if the primary proton injection spectrum is hard, the shape of the MeV-GeV synchrotron bump in the spectrum depends slightly on the details of the proton injection spectrum, because the shape of the e$^+$e$^-$ pair spectrum (shown in  the upper panel of Fig. \ref{fig:without_pion}) is determined by the radiative cooling effects, rather than by the details of the e$^+$e$^-$ injection spectrum (shown by the dashed line in the upper panel of the Figure).  The hard e$^+$e$^-$ pair spectrum below $E\sim 10^8$~eV is explained by the dominance of the Coulomb loss in the densest innermost part of the stellar wind. The $E^{-2}$ type spectrum between $10^8$~eV and $10^{10}$~eV is explained by the dominance of the synchrotron loss in this energy range. The hardening of the spectrum in the range $10^{10}$~eV$\,<E\,<\,10^{12}$~eV is explained by the dominance of the IC energy loss proceeding in the Klein-Nishina regime. Above $10^{12}$~eV, the IC cooling becomes less efficient than the synchrotron cooling, which leads to the softening of the spectrum. 

The electron energy at which the synchrotron loss takes over the Klein-Nishina energy loss can be estimated by comparing the synchrotron cooling time (\ref{tsynch}) to the inverse Compton cooling time (\ref{tkn}):
\begin{equation}
E_{KN-S}\simeq 10^{13}\left[\frac{B_0}{1\mbox{ G}}\right]^{-1}\mbox{ eV}
\end{equation}
(where $B_0$ is the magnetic field in the innermost part of the stellar wind and we assume the radial profile $B\sim D^{-1}$). The synchrotron emission by electrons of such energies is produced in the energy band above 
\begin{equation}
\epsilon>\epsilon_{\rm synch-bump}\simeq 3\left[\frac{D}{R_*}\right]^{-1}\left[\frac{B_0}{1\mbox{ G}}\right]^{-1}\mbox{ MeV, }
\end{equation}
 resulting in a broad "bump" in the synchrotron spectrum above $\epsilon_{KN-S}$. 

The comparison of synchrotron, inverse Compton and Bremsstrahlung emission spectra from the "cooling-shaped" electron spectrum to the \lsi\ multi-wavelength data \cite{chernyakova06a}, done in the lower panel of Fig. \ref{fig:without_pion}, shows that the model in which all the e$^+$e$^-$ pairs are initially injected at very high energies (above $100$~TeV) provides a good fit to the broad-band spectrum of the source. To produce this figure, we have taken into account the attenuation of the spectrum of the \gr\ emission due to the $\gamma\gamma$ pair production in the photon field of the Be star, assuming that the \gr s escape almost radially from the emission region. This corresponds to the situation of the compact object close to the inferior conjunction of the orbit, where the effects of the absorption of the \gr\ flux are minimized. 

\begin{figure} 
\centerline{\includegraphics[width=0.9\linewidth]{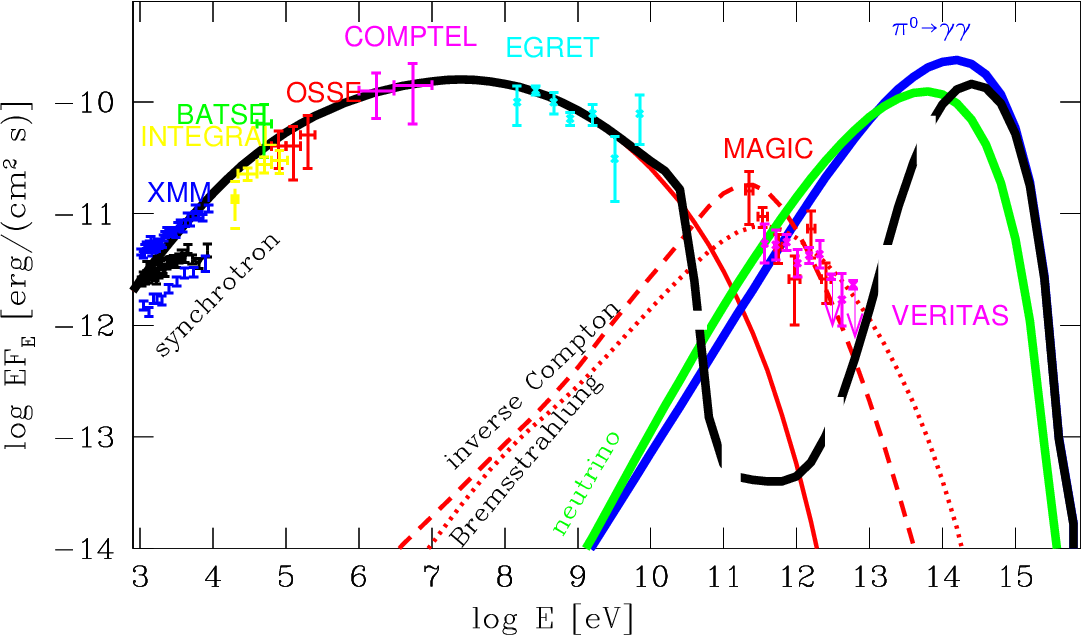}}
\caption{Broad-band spectrum of emission from pp interactions, calculated assuming the same parameters as in Fig. \ref{fig:without_pion}, but considering the possibility of strong neutrino emission from the source. The notations for the \gr\ spectrum are the same as in the lower panel of Fig. \ref{fig:without_pion}. The green thick solid line shows the neutrino spectrum. Blue thick solid line shows the $\pi^0$ decay contribution to the \gr\ spectrum. The black thick line shows the overall broad-band model spectrum. The dashed part of the line shows the part of the spectrum attenuated by $\gamma\gamma$ pair production.}
\label{fig:gamma0} 
\end{figure}

Fig. \ref{fig:gamma0} shows an opposite situation in which the \gr\ spectrum strongly absorbed in the energy band 0.1-1~TeV (long-dashed thick solid line) due to the pair production. This situation can occur when the emission comes from the densest innermost part of the stellar wind in the vicinity of the Be star. The real value of the average $\tau_{\gamma\gamma}$ cannot be estimated unless the details of the 3-dimensional geometry of the emission region and the relative orientations of the extended emission region, of the Be star and of the observer are known. Taking into account this uncertainty, we choose for $\tau_{\gamma\gamma}$ the minimal value 
which ensures the non violation of the flux upper bound at energies above $\sim (several)$~TeV from VERITAS observations of the source \cite{veritas}. 

Fig. \ref{fig:gamma0} also shows the \gr\ emission spectrum from neutral pion decays (blue thick solid line) and the neutrino spectrum (green thick solid line).  
The observation of the  MeV-GeV band synchrotron emission from the secondary e$^+$e$^-$ pairs enables to constrain of the possible neutrino emission as explained previously.
Contrary to the emission from e$^+$e$^-$ pairs, neutrino and $\pi^0$ decay emission is anisotropic. In principle, the anisotropy could lead to a boosting of the neutrino and $\pi^0$ decay components. 
However, if the orbital plane is not aligned with the line of sight, the neutrinos and \gr s from $\pi^0$ decay should be emitted into a cone with rather wide opening angle, to be observable (the opening angle of the cone should be larger than the inclination angle of the binary orbit, see Fig. \ref{fig:ellipse}). This means that the flux is at most moderately enhanced by the anisotropy effect. At the same time, if the inclination of the orbit is larger than the opening angle of the neutrino and $\pi^0$ decay \gr\ emission cone, the flux in these emission components could be negligible.

\begin{figure} 
\centerline{\includegraphics[width=0.9\linewidth]{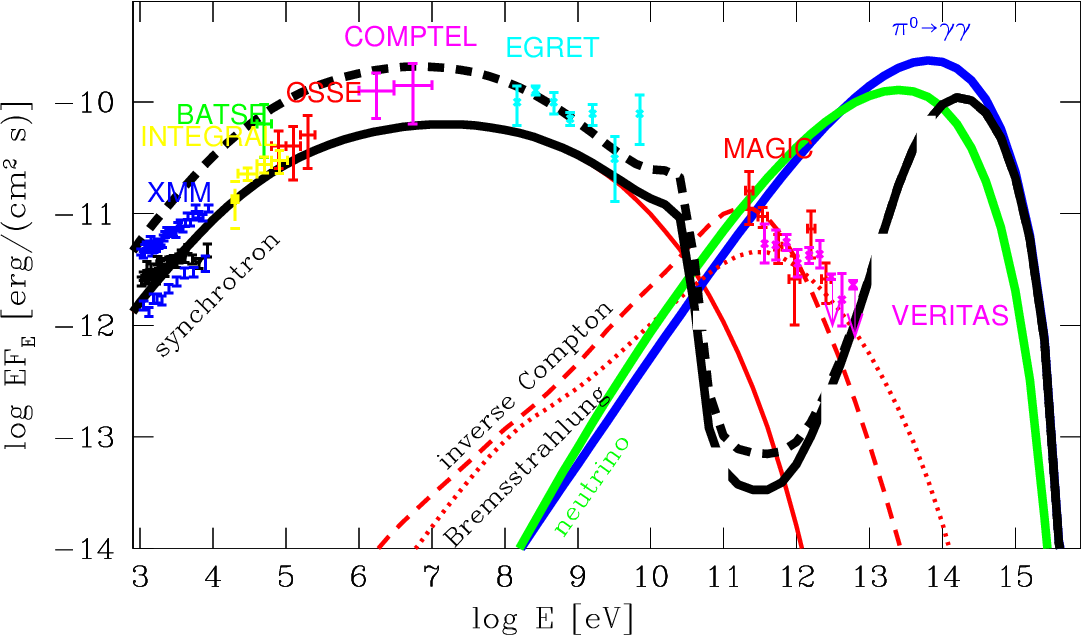}}
\caption{Same as in Fig. 3, but for $\Gamma_p=1$. The thick short dashed line shows the overall spectrum calculated taking into account the tertiary ${\rm e}^+{\rm e}^-$ pairs from $\gamma\gamma$ interactions.}
\label{fig:gamma1} 
\end{figure}

\begin{figure} 
\centerline{\includegraphics[width=0.9\linewidth]{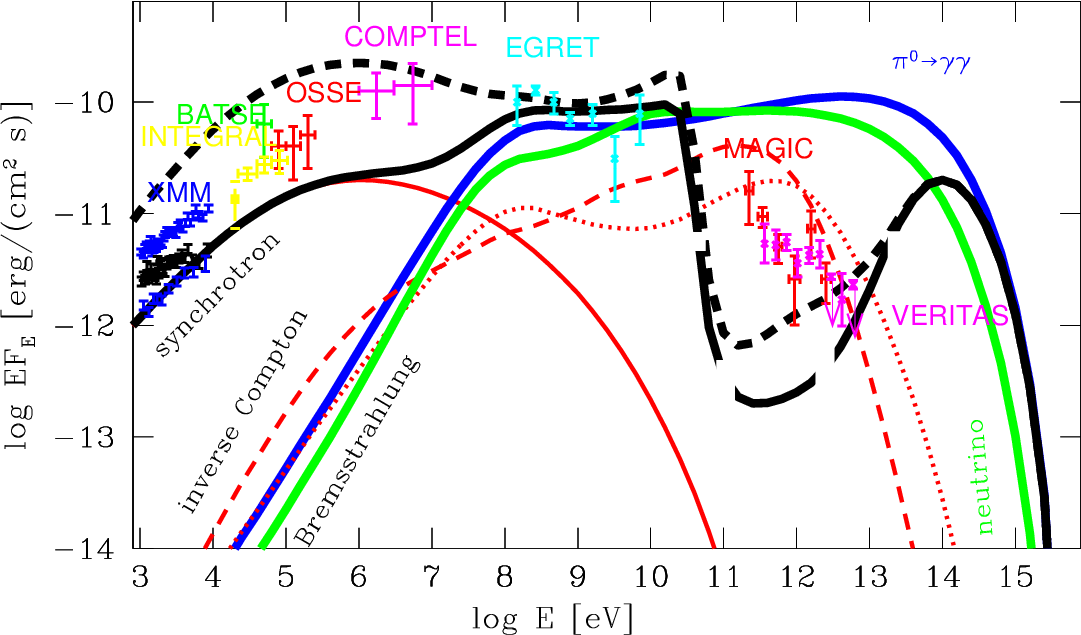}}
\caption{Same as in Fig. 4 but for the proton injection spectrum $\Gamma_p=2$. The magnetic field is assumed to be $B_0=0.5$~G. }
\label{fig:gamma2} 
\end{figure}

Figs. \ref{fig:gamma1} and \ref{fig:gamma2} show the results of calculation of the broad-band emission spectrum from pp interactions obtained assuming softer spectra for the primary protons, with $\Gamma_p=1$ and $\Gamma_p=2$, but keeping the same cut-off energy, $E_{\rm cut}=10^{15}$~eV.
In these figures, we illustrate the effect of an additional uncertainty in the calculation of the broad-band spectrum of the source, related to the uncertainty of the cascade contribution to the observed source spectrum. The problem is that the shape of the electron spectrum below 10~TeV is  modified by the injection of ${\rm e}^+{\rm e}^-$ pairs via the process of absorption of the \gr s on the UV photon background. In case the synchrotron radiation dominates the energy loss of the "tertiary" electrons, the synchrotron emission from the cascade ${\rm e}^+{\rm e}^-$ pairs will further increases the height of the MeV-GeV synchrotron bump in the spectrum. In case inverse Compton dominates the energy loss of the cascade electrons, the tertiary pairs will emit mostly in the 1-10~GeV energy band. The overall flux of the cascade contribution is determined by the average optical depth of the extended source with respect to the pair production. The thick short dashed lines in Figs. \ref{fig:gamma1} and \ref{fig:gamma2} show the calculated modification of the broad-band spectrum of the source by the cascade.


\section{Estimate of the number of neutrino events for ICECUBE}
\label{sec:icecube}

In the previous section we have shown that the hadronic model of activity of \gr -loud binaries could provide a good fit to the broad-band spectra of these sources. Within this model, the observed MeV-GeV bump in the spectral energy distribution is explained by the synchrotron emission from the secondary e$^+$e$^-$ pairs produced either in pp interactions or in result of $\gamma\gamma$ pair production process. 

A direct test of the hadronic model would be the detection of neutrino flux from the known \gr -loud binaries. In principle, the expected neutrino flux from a given source can be estimated from the known \gr\ flux. In the simplest case when the source is transparent to the \gr s, the estimate is straightforward, because the normalization and spectrum of the \gr\ flux produced by the neutral pion decays in the source directly gives an estimate of the spectrum and normalization of the neutrino flux. However, modeling of the broad-band spectra of emission from \gr -loud binaries, described in the previous section, shows that this simple possibility, most probably, does not hold. First, the spectrum of \gr\ emission in the TeV band is affected by the $\gamma\gamma$ pair production, so that neither the overall flux, nor the spectral index of the \gr\ emission are related to the ones of the neutrino spectrum. Next, a sizable contribution to the spectrum of \gr\ emission in the TeV band can be given by the IC emission from the secondary ${\rm e}^+{\rm e}^-$ pairs produced in pp interactions. This further modifies the flux and the spectrum of TeV \gr\ emission.

Thus, the information on the properties of the TeV \gr\ emission from a \gr -loud binary can not be used to estimate the neutrino flux from the source. Surprisingly, the source flux in the MeV-GeV, rather than TeV energy range can be used for the neutrino flux prediction. This possibility arises because within the hadronic model of activity, the MeV-GeV bump in the spectral energy distribution is produced by the emission from the secondary ${\rm e}^+{\rm e}^-$ pairs from the pp interactions. Contrary to the TeV flux, the MeV-GeV flux from the \gr -loud binary is not affected by the pair production and can be used to estimate the total energy output from the pp interactions in the source. The only uncertainty of such an estimate is that the synchrotron emission from the secondary ${\rm e}^+{\rm e}^-$ pairs in the MeV-GeV energy band is, most probably, isotropic, while the  neutrino and $\pi^0$ decay \gr\ emissions are not. As explained in the previous section, the observed source flux in the MeV-GeV band gives, in fact, an upper limit on the possible neutrino flux from the source.   

Although the total  power of the neutrino emission can be estimated from the MeV-GeV luminosity of the \gr -loud binary, the modelling of the broad-band emission spectrum of the source gives only mild constraints on the neutrino emission spectrum: acceptable models of the broad-band spectra can be found assuming initial proton injection spectra ranging from a $E^{-2}$ power law to an almost monochromatic injection spectrum. 

Below we estimate the maximal neutrino event rate in ICECUBE for a particular example of LSI +61 303, taking into account the uncertainty of the spectrum of neutrino emission from the source. 
In the most optimistic case scenario, when the neutrino flux is at the level of the observed MeV-GeV flux from the source, we also investigate the possibility to measure the parameters of the neutrino spectrum (and, therefore, of the primary proton spectrum), such as the spectral slope and/or the cut-off energy.

The rate of neutrino-induced muon events from a point source at declination $\delta$ with differential neutrino spectrum ${\rm{d}}\Phi_\nu(E_\nu)/{\rm{d}E_\nu}$ 
\begin{eqnarray}
\label{eq:effareas}
{\cal R}_\mu&=& \int {\rm{d}}E_\nu A^{\rm{eff}}_\nu(E_\nu,\delta) \frac{{\rm{d}}\Phi_\nu(E_\nu)}{{\rm{d}}E_\nu} \label{eq:nevnu} \\
&=& \int {\rm{d}}E_\mu A^{\rm{eff}}_\mu(E_\mu) \frac{{\rm{d}}\Phi_\mu(E_\mu,\delta)}{{\rm{d}}E_\mu}. \label{eq:nevmu}
\end{eqnarray}
where $A_{\rm{eff}}(E_\nu,\delta)$ and $A^{\rm{eff}}_\mu(E_\mu)$ are respectively the neutrino and the muon effective areas of the detector. The neutrino effective area, often provided in papers in contrast to the muon effective area, folds the detector efficiency and the neutrino propagation effects. Therefore Eq.~(\ref{eq:nevnu}) is convenient for a direct calculation of the number of detected neutrino-induced muon events in a specified neutrino energy range.

However, the neutrino energy is not an experimental observable, contrary to the reconstructed muon energy. If we are interested in understanding whether the different neutrino spectra can be experimentally differentiated, {\it i.e.} we want to assess the spectral response of the detector for each model, we should use instead Eq.~(\ref{eq:nevmu}). This cannot be done without further derivation of the muon effective area and knowledge of the details of the detector, namely the angular resolution (of the order of 1$^\circ$~\cite{sendai} for ICECUBE in its final 80 string configuration (IC80) and precisely accounted for in the following) and the energy resolution of the order $\sigma_{\log{E}}\approx 0.3$~\cite{dima,composite-ribordy}.

The differential neutrino induced muon spectrum at the detector, for a given source at location $\delta$, ${\rm{d}}\Phi_\mu(E_\mu,\delta)/{\rm{d}}E_\mu$, is  obtained after propagation of neutrinos up to the interaction point and further propagation of the muons to the detector. This can be sketched as follows: 
\begin{equation}
\label{scheme}
E_\nu \stackrel{y_{\rm{CC}}(E_\nu)}{\longrightarrow} E_{\mu_0} \stackrel{p_{\rm{det}}(E_{\mu_0},E_\mu)}{\longrightarrow} E_\mu  \stackrel{p_{\sigma_E}(E_\mu, E_\mu^{\rm{rec}})}{\longrightarrow} E_\mu^{\rm{rec}}.
\end{equation}
Here $E_{\mu_0}(E_\nu)$ is the muon energy at the interaction vertex, given by $E_{\mu_0}(E_\nu) = (1 - y_{\rm{CC}}(E_\nu))E_\nu$, where $y_{\rm{CC}}(E_\nu)$ is the mean interaction inelasticity, taken from~\cite{gqrs} (calculations below will adopt a simplifying hypothesis, considering an average inelasticity). $p_{\rm{det}}(E_\mu,E_\nu)$ is the probability density for a muon produced with energy $E_{\mu_0}(E_\nu)$ to reach the detector with energy $E_\mu$. for a given distance between the interaction vertex and the detector, $E_\mu$ is calculated from $E_{\mu_0}(E_\nu)$ by integrating the muon energy loss rate $dE_\mu/dX=-(a+bE_\mu)$, where the $a,\,b$ are the energy independent standard rock coefficients from Ref. \cite{mmc} and $X$ is the grammage traversed by the muon on its way to the detector, 
\begin{eqnarray} 
p_{\rm{det}}(E_\mu,E_\nu) &=& \frac{-{\rm{d}}X(E_\mu,E_{\mu_0}(E_\nu))/{\rm{d}}E_\mu}{R_\mu(E_{\mu_0}(E_\nu))}
 \\
&=&  \frac{1}{\ln{(1+E_{\mu_0}(E_\nu)/\epsilon)}}\frac{1}{(E_\mu+\epsilon)},
\end{eqnarray} 
where $\epsilon=a/b$.
The function $p_{\rm det}(E_\mu,E_\nu)$ is set to zero outside of the interval $E_\mu \in [0,E_{\mu_0}(E_\nu)]$. $X(E_\mu,E_{\mu_0})$ is the grammage function for a muon of initial energy $E_{\mu_0}$ and final energy $E_\mu$ and $R_\mu(E_{\mu_0}) \equiv X(0,E_{\mu_0})$ is its range.
Finally, the function $p_{\sigma_E}(E_\mu, E_\mu^{\rm{rec}})$, which enters the final stage of calculation sketched in (\ref{scheme}) characterizes the energy resolution of the detector and is discussed below.

The differential muon flux at the detector, which enters equ. (\ref{eq:nevmu}) can be calculated via the propagation of the muon flux from the interaction vertex as
\begin{eqnarray}
\label{Fmu}
\frac{{\rm{d}}\Phi_\mu(E_\mu,\delta)}{{\rm{d}}E_\mu} &=& \int {\rm{d}}E_\nu p_{\rm{det}}(E_\mu,E_\nu) \nonumber \\
&\times& p_{\rm{tr}}(E_\nu,\delta)p_{\rm{int}}(E_\nu) \frac{{\rm{d}}\Phi_\nu(E_\nu)}{{\rm{d}}E_\nu}.
\end{eqnarray}
where $p_{\rm int}(E_\nu)=N_A \sigma_{\rm{CC}}(E_\nu) R_\mu(E_{\mu_0})$ is the neutrino interaction probability in the vicinity of the detector (potentially producing a muon within the reach of the detector) and $p_{\rm tr}(E_\nu,\delta)=\exp{(-N_A \sigma(E_\nu) X(\delta))}$ is the earth transmission probability of a neutrino arriving from declination $\delta$ ({\it i.e.} after crossing a grammage $X(\delta)$). $N_A$ is the Avogadro number, $\sigma_{\rm CC}$ and $\sigma$ are respectively the charged current and the total muon neutrino cross sections, taken from \cite{mrsg}.

The separability feature of the kernel $p_{\rm{det}}(E_\mu,E_\nu)$, together with the implicit assumption of the absence of muon propagation fluctuation allows us to analytically extract $A_\mu^{\rm{eff}}$ by substituting Eq. (\ref{Fmu}) into (\ref{eq:nevnu}),~(\ref{eq:nevmu}):
\begin{eqnarray}
\label{Amu}
A^{\rm{eff}}_\mu(E_\mu) =\left.\frac{ \frac{\displaystyle d}{\displaystyle dE_\nu}\Big(
\frac{\displaystyle A^{\rm eff}_\nu(E_\nu)}{\displaystyle p_{\rm det}(E_\mu,E_\nu)p_{\rm int}(E_\nu)}\Big)}
{dE_{\mu_0}(E_\nu)/dE_\nu} 
\right|_{E_\nu = E_{\mu_0}^{-1}(E_\mu)}
\end{eqnarray}
The neutrino effective area $A^{\rm{eff}}_\nu (E_\nu)$ without absorption is extracted from the neutrino effective area averaged over the whole northern hemisphere, $\langle A^{\rm{eff}}_\nu (E_\nu,\delta) \rangle_\delta$  from~\cite{sendai}, using the relation
\begin{equation} 
A_\nu^{\rm{eff}}(E_\nu) = 
\frac{\langle A^{\rm{eff}}_\nu (E_\nu,\delta) \rangle_\delta}{\int_{-1}^0 p_{\rm{tr}}(E_\nu,\theta-\pi/2)\, d(\cos{\theta})}
\end{equation}
The transmission probability $p_{\rm{tr}}(E_\nu,\delta)$ which appears in the denominator (numerically computed using the Preliminary Earth Model~\cite{prem}) strongly decreases with energy. Therefore, to estimate the neutrino effective area at the declination of \lsi, the effect of earth absorption is accounted for according to
\begin{equation}
A_\nu^{\rm{eff}}(E_\nu,\delta_{\rm{LSI+61\,303}}) =  p_{\rm{tr}}(E_\nu,\delta_{\rm{LSI+61\,303}}) A_\nu^{\rm{eff}}(E_\nu).
\end{equation}

It should be mentioned that $A_\nu^{\rm{eff}}$ is approximately, at the modest energy considered here, the effective area at zero declination, assuming an isotropic detector response and analyses uniform down to the horizon (of course, analyses proceed non completely uniformly w.t.r. the declination to account for the difficult background rejection near the horizon. Nevertheless, lately, as can be seen e.g. in~\cite{amanda}, there is a remarkable uniformity of the sensitivity, down to very close to the horizon).

\begin{figure} 
\centerline{\includegraphics[width=\linewidth]{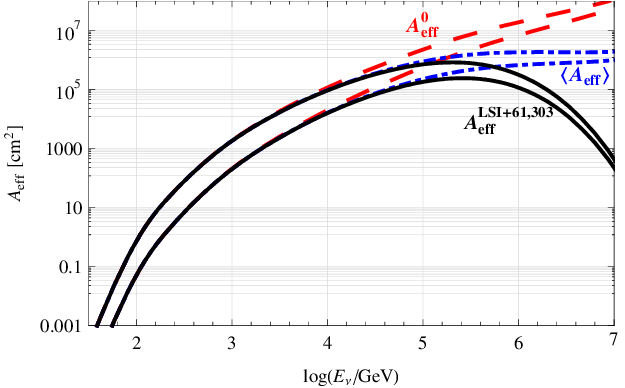}}
\centerline{\includegraphics[width=\linewidth]{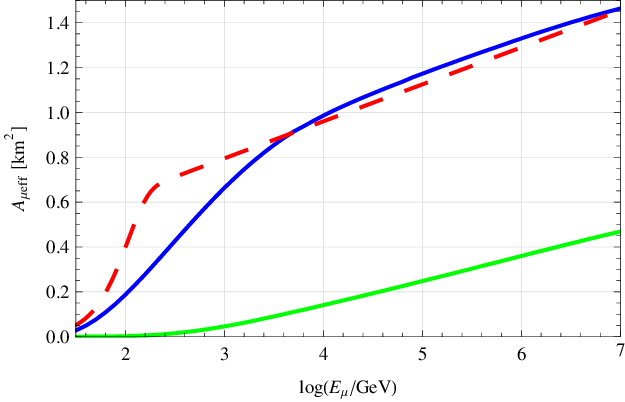}}
\caption{Top: Comparison of different neutrino effective areas. Red dashed line showns the effective area $A_\nu^{\rm{eff}}$ with no absoption. Blue dot-dashed line shows the effective area after averaging over the declinations $\delta$ for an isotropic neutrino flux $\langle A_\nu^{\rm{eff}}\rangle$~\cite{sendai}. Curves running in parallel are for IC22 and IC80. Black solid line shows the effective area for \lsi. Bottom: Full lines are the muon effective area calculated with~(\ref{Amu}) for IC22 and IC80. The dashed curve is the muon effective area derived in~\cite{halzen-newphys}.}
\label{fig:nu_spectranu_aeff} 
\end{figure}

The effective areas are shown for IC22 (the ICECUBE 22 string configuration) and IC80, in Fig.~\ref{fig:nu_spectranu_aeff}. In the lower panel, the ICECUBE muon effective area derived in~\cite{halzen-newphys} shows a remarkable agreement above $\sim$1~TeV, while the behavior at lower energy is rather different: our calculation exhibits a soft exponential decrease below $\sim$1 TeV contrasting to the sharp cutoff at $\sim$100 GeV.
Considering the upper panel, we notice the fast decrease of the neutrino effective area at LSI+61 303 declination above $\sim$300~TeV. This is due to the fact that the neutrino trajectories towards ICECUBE are slightly passing through the high density earth core. This leads to a significant suppression of the neutrino flux above this energy (the observation of PSR B1259-63 and LS 5039 with the Km3Net\cite{km3net} would not suffer this limitation and thus offer the possibility to probe spectral features up to the highest cutoff energies).

Using the analytical muon effective area (\ref{Amu}), the diffuse atmospheric neutrino background rate and the point source signal event rate as a function of the chosen reconstructed muon threshold energy $E_{\rm thr}$ can be calculated,  
\begin{eqnarray}
&&{\cal R}_{\rm{b}}(E_\mu^{\rm{rec}} > E_{\rm{thr}}, \psi,\delta) 
 =  \int {\rm{d}}E_\mu \,\frac{{\rm{d}}\Phi^{\rm{atm}}_\mu(E_\mu,\delta)}{{\rm{d}}E_\mu}  \nonumber\\
&& \times \, f_{\rm{b}}(\psi) \, A_\mu^{\rm{eff}}(E_\mu) \omega_{\sigma_E}(E_\mu,E_{\rm{thr}}) \label{equ:nb} \\
\nonumber && {\cal R}_{\rm{s}}(E_\mu^{\rm{rec}} > E_{\rm{thr}}, \psi,\delta) 
 =  \int {\rm{d}}E_\mu \,\frac{{\rm{d}}\Phi^{\rm{model}}_\mu(E_\mu,\delta)}{{\rm{d}}E_\mu}  \nonumber\\
&& \times \, f_{\rm{s}}(\psi,E_\mu) \, A_\mu^{\rm{eff}}(E_\mu) \omega_{\sigma_E}(E_\mu,E_{\rm{thr}}) \label{equ:ns}
\end{eqnarray}
$f_{\rm{b}}(\psi)=\pi\psi^2$ is the solid angle subtended within a radius $\psi$ for the diffuse atmospheric neutrino background and $f_{\rm{s}}(\psi,E_\nu)$ is the fraction of signal events with energy $E_\nu$ reconstructed within an angle $\psi$ from the source:
\begin{equation}
f_{\rm{s}}(\psi,E_\nu) = \sum_{i=1,2} a_i\left(1-\exp\left[-\frac{{\psi^2}}{2\sigma_{i,\rm{\psi}}^2(E_\nu)}\right]\right),
\end{equation}
where the point spread function (PSF) $\sigma_{i,\Psi}(E_\nu)=\sqrt{\sigma_{i,\Psi_{\mu\mu_{\rm{rec}}}}^2 + \Theta_{\nu\mu}^2(E_\nu)}$ decomposes into two distinct PSF contributions: the energy-dependent neutrino to muon kinematics $\Theta_{\nu\mu}(E_\nu) = 0.54^\circ / \sqrt{E_\nu/{\rm TeV}}$ and the muon reconstruction resolution $\sigma_{i,\Psi_{\mu\mu_{\rm{rec}}}}$.
For our analysis we extracted the parameters $\sigma_{i,\Psi_{\mu\mu_{\rm{rec}}}}$ and $a_i$  from a fit to the IC22 and IC80 PSF found in~\cite{sendai} (for IC80, we obtain: $a_1^{\rm{IC80}}=0.44,\,\sigma_{1,\rm{\psi}}^{\rm{IC80}}=0.47^\circ,\,a_2^{\rm{IC80}}=0.51,\,\sigma_{2,\rm{\psi}}^{\rm{IC80}}=1.12^\circ$. E.g. at 10 TeV, 50\% of the neutrinos coming from a specified source are reconstructed within 0.9$^\circ$ for IC80).
In the following, the results will be presented using the optimized $\psi=1.3^\circ$ for IC80.

Above, in eqs~\ref{equ:nb} and~\ref{equ:ns}, the energy argument for $f_{\rm{s}}$ is set to $E_\nu=E_\mu$. This is a conservative statement because the muon energy is smaller than the neutrino energy, we have $f_{\rm{s}}(\psi,E_\mu) < f_{\rm{s}}(\psi,E_\nu(E_\mu))$. 

The weight factor used to calculate the number of events above some reconstructed energy:
\begin{eqnarray}
\omega_{\sigma_E} (E_\mu,E_{\rm{thr}}) &=& \int_{E_{\rm{thr}}}^\infty p_{\sigma_E} (E^{\rm{rec}}_\mu, E_\mu) {\rm{d}}E_\mu^{\rm{rec}} \\ 
~ &=& \frac{1}{2} {\rm{erfc}}(\frac{\log{E_{\rm{thr}}/E_\nu} } { \ln{10}\sqrt{2} \sigma_E} ).
\end{eqnarray}
follows from an assumed gaussian energy resolution distribution function in $\Delta\log{E}$,
$p_{\sigma_E}(E_\mu^{\rm{rec}},E_\mu)=p_{\sigma_E}(\log{E_\mu^{\rm{rec}}}-\log{E_\mu})$, with $\sigma_E=0.3$.

In our calculations, the muon propagation was implemented through $-({\rm{d}}E/{\rm{d}}X) = a+bE$, with the energy-independent parameters $a$ and $b$ for standard rock (adequate, given the source location and the proximity of ICECUBE to the bedrock), in order to derive the muon energy distribution at the detector~\cite{mmc}. Therefore, fluctuations of muon energy due to the stochastic nature of high energy muon energy losses which implies muon energy distributions, were not accounted for~\cite{stanevgaisser}. A more refined treatment, which requires numerical methods, proceed in integrating over the distance the probability $F(E_{\mu_0},E_\mu,l)$~\cite{liparistanev,halzen-newphys} for a muon generated at distance $l$ with energy $E_{\mu_0}$ to reach the detector with energy $E_\mu$.
Our simplified treatment has nevertheless the advantage of transparently combining the latest published neutrino effective area with some phenomenological assumptions for the detector muon effective area, detector response, muon propagation and neutrino interaction, resulting in a simple analytical formula for the rates, thus avoiding more generic integro-differential forms.

We choose the angular-dependent parametrization from Volkova (prompt + conventional neutrino fluxes) for the atmospheric neutrino background model~\cite{volkova}. 
This corresponds to 7.7 detected muons when using the neutrino effective area and per year with IC80 in the 1.3$^\circ$ circular bin at \lsi\ declination. Conversely, calculating this number with the analytical muon effective area~\ref{Amu} yields 8.8 detected muons. This 15\% difference may come from the various simplifications and is probably dominated by the delta function differential cross section approximation (at low energies, the average range of the neutrino induced muons is larger). At \lsi\ declinations, the atmospheric neutrino zenith angle to be considered is $\theta=\delta-\arcsin{(2\sin^2{\delta}-1)}=29^\circ$ ($\theta=0^\circ$ is for vertical incidence).

\begin{figure}
\centerline{\includegraphics[width=\linewidth]{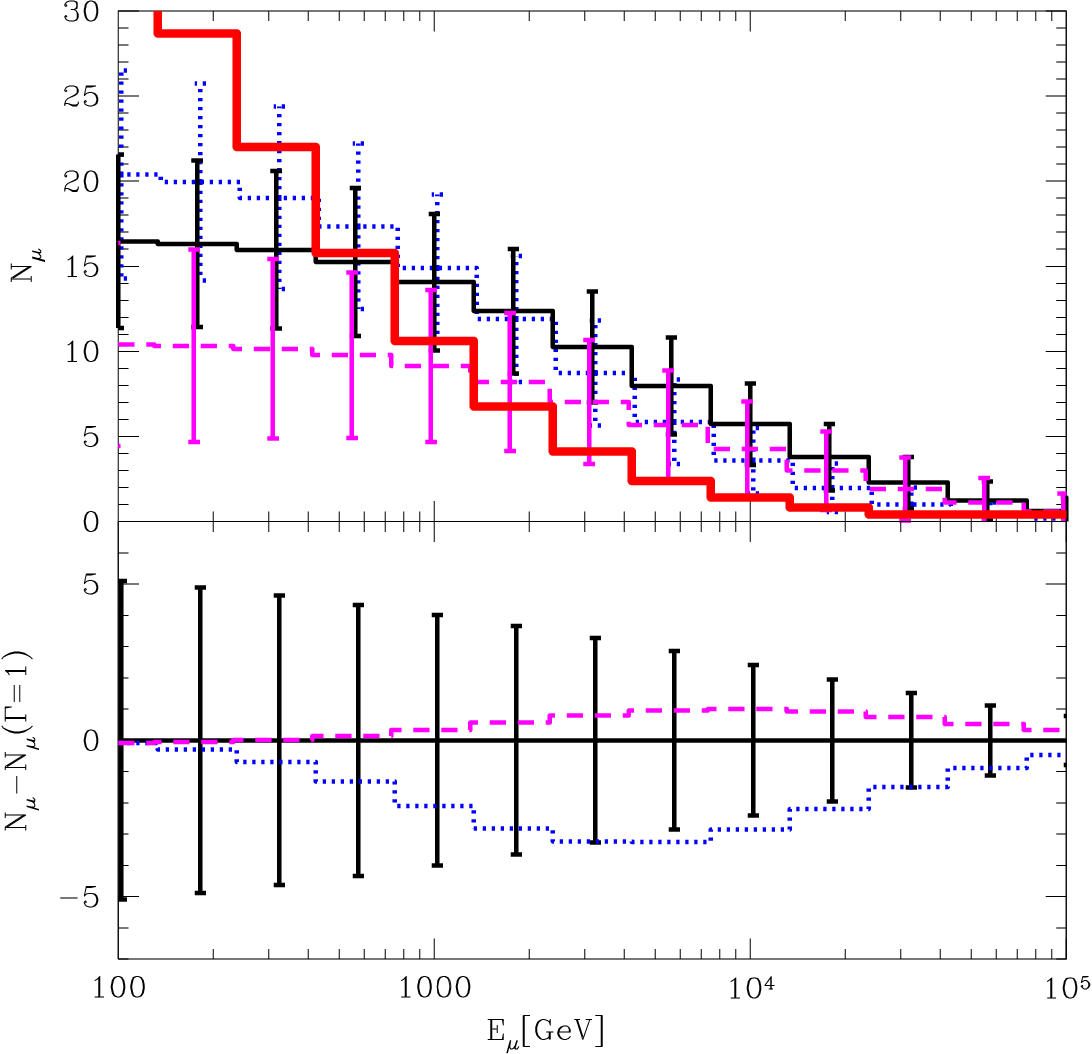}}
\caption{Background-subtracted cumulative muon spectra $N(E_{\mu, \rm thr})$, expected after the 3-year ICECUBE exposure for the three model neutrino spectra of \lsi, discussed in the previous sections (error bars of signal muon spectra are the sum in quadrature of statistical errors of signal + atmospheric neutrino background). The blue dotted / black solid / magenta dashed line respectively show the spectra corresponding to proton injection spectra with spectral indices $\Gamma=2$, $\Gamma=1$ and $\Gamma=0$. The red thick solid line shows the $5\sigma$ excess above the atmospheric neutrino background (the "discovery threshold").}
\label{fig:nu_spectra} 
\end{figure}

The results, in terms of the atmospheric background-subtracted muon spectra for the baseline neutrino flux models discussed in the previous sections, are presented in Fig.~\ref{fig:nu_spectra}. The exposure time is taken to be 3 years of running the full ICECUBE array. For comparison, we also show by the red solid thick line the $5\sigma$ level above the atmospheric background (extracted from binomial statistics, where the trial factor is the total expected number of atmospheric  neutrino $n(E)$ in a 2$\psi$ width declination band reconstructed with energy $>E$ and the probabiblity of success is $p=\pi\psi^2/2\pi\int_{\delta-\psi}^{\delta+\psi}{\rm{d}}\cos{\theta}$ in the declination band, $n_{5\sigma}(E)=np+5\sqrt{np(1-p)}$. Note however that an unbinned calculation slightly lower the signal requirements \cite{simul}).

We calculate the number of muon events as a function of the energy threshold for the three neutrino spectra discussed in the previous section. The blue dotted line show the detected muon spectrum for the model with the proton injection spectrum with spectral index $\Gamma_p=2$, the solid black line corresponds to the proton injection spectrum with $\Gamma_p=1$ and cut-off at the same energy, while the magenta dashed line corresponds to $\Gamma_p=0$. In all three cases the cut-off energy is assumed to be $E_{\rm cut}=1$~PeV.
 
One can see that the uncertainty of the proton injection spectra results in a factor of $\sim 2$ uncertainty in predictions of the number of neutrinos detectable in ICECUBE. Assuming that the neutrino flux saturates the upper limit imposed by the observed \gr\ flux in the MeV-GeV energy band, one finds that the neutrino source should be detectable at $5\sigma$ level in the energy band $1-10$~TeV in roughly one year, if the proton injection spectrum is not too hard. 

Inspecting the muon spectra shown in the upper panel of Fig. \ref{fig:nu_spectra}, one sees that softed proton injection spectrum results in a slight excess of muon events at lower energies. If the overall normalization of the neutrino flux were known, the measurement of the muon event spectrum would allow to constrain the spectrum of the primary protons in the source. However, taking into account the uncertainty of the overall normalization of neutrino flux introduced by the uncertainty of the anisotropy pattern of neutrino emission, one finds that the statistics of the signal will in the best case only allow for marginal estimate of the neutrino spectrum parameters. This is illustrated in the lower panel of Fig. \ref{fig:nu_spectra}, where a comparison of the shapes of the muon spectra is shown. If one assumes the same total number of muon events, the difference in the spectra for the three models mostly remain within $\sim 1\sigma$ error bars over the entire energy range.

The neutrino effective area in the direction of \lsi\ peaks at the energy $E_\nu\sim 100$~TeV (see Fig. \ref{fig:nu_spectranu_aeff}. Taking into account the typical inelasticity of pp interactions, one finds that neutrinos of this energy are injected by the primary protons with energies  $E_p\sim 1$~PeV. This means that assuming the cut-off in the proton spectrum at the energy $E_{\rm cut}\sim 1$~PeV, one maximizes the number of neutrino events detectable with ICECUBE. If the cut-off in the proton spectrum is much above 1~PeV, the neutrino signal will be lost because of the opacity of the Earth. Otherwise, if the cut-off in the proton spectrum is much below 1~PeV, the signal is lost because of the transparency of the detector for the neutrinos and because of the high atmospheric neutrino background. Thus, the typical numbers of neutrino-induced events in the ICECUBE, shown in Fig. \ref{fig:nu_spectra} should be considered as the upper bound on the possible neutrino signal from \lsi.

\section{Conclusions}

We have estimated the neutrino flux from GRLBs expected within the hadronic model of activity of these sources. Within such a model, the measured spectral characteristics of \gr\ emission from the source in the TeV energy band are not directly related to the spectral characteristics of the neutrino emission, because of absorption of the TeV \gr s on the thermal photon background produced by the massive star in the system. The uncertainty of the calculation of the attenuation of the TeV \gr\ flux introduces a large uncertainty to the estimate of the neutrino flux based on the measured TeV \gr\ flux. 

Taking this uncertainty into account, we have adopted a different approach for the estimate of the neutrino flux from a GRLB. Namely, we have noted that the energy output of proton-proton interactions, and hence the neutrino flux, can be constrained by the broad-band spectrum of the source. The idea is that the ${\rm e}^+{\rm e}^-$ pairs, produced in the decays of charged pions, release their energy in the form of the synchrotron, IC and Bremsstrahlung emission, which contributes to the observed broad-band SED of the source. The requirement that the power of electromagnetic emission from the secondary ${\rm e}^+{\rm e}^-$ pairs in different energy bands is not higher than the observed source luminosity imposes a constraint on the neutrino luminosity of the GRLB.

We have worked out the numerical model of the broad-band emission from a GRLB, assuming that pp interactions take place in the vicinity of the bright massive star, in the innermost and densest part of the stellar wind. We have shown that the interactions of high energy protons (with energies reaching $\sim$PeV) in the dense stellar wind can explain the observed broad-band emission, with the MeV-GeV "bump" of the SED being due to the synchrotron emission from the secondary ${\rm e}^+{\rm e}^-$ pairs produced either in the decay of charged pions or resulting from the absorption of 10~GeV-TeV \gr s on the thermal photons from the massive star (see Figs. 1-4). 

Although the observed bolometric luminosity of the source (i.e. the height of the MeV-GeV bump of the SED) constrains the overall neutrino luminosity, the shape of the neutrino spectrum and the overall normalization of the neutrino flux are only mildly constrained by the multi-wavelength data. The problem is that the properties of the broad-band emission from the secondary ${\rm e}^+{\rm e}^-$ pairs are mostly determined by the effects of the radiative cooling and the escape of the pairs from the source, rather than by the initial injection spectrum of the pairs. This means that quite different injection spectra of the primary high energy protons (and hence of the neutrinos) can result in approximately the same broad-band SEDs, as is clear from Figs. 1-4. In addition, the anisotropy pattern of neutrino emission is expected to differ from that of the synchrotron emission from the ${\rm e}^+{\rm e}^-$ pairs. This results in an uncertainty  of the anisotropy-related suppression of the neutrino flux compared to the MeV-GeV synchrotron flux. 

Taking into account these uncertainties of the neutrino emission spectrum, we have estimated the expected number of neutrinos which will be detected by ICECUBE, assuming that the neutrino flux saturates the upper bound imposed by the observed \gr\ flux in the MeV-GeV energy band. 
Considering the particular example of \lsi, we have found that if the spectrum of high energy protons in the source extends to the PeV energies, the source would be readily detectable within roughly one year of exposure with ICECUBE.  

We have also explored the potential of the full ICECUBE detector for the measurement of the spectral characteristics of the neutrino signal from \lsi. We find that in the case when the neutrino flux is at the level of the upper bound imposed by the observed MeV-GeV \gr\ flux, an exposure time longer than 3 years will be required to reliably constrain the spectral index of the primary high energy proton spectrum  via observations of neutrino signal in ICECUBE. The mere fact of a possible spectral characterization of neutrino sources within the lifetime of ICECUBE represents however a most exciting prospect for the future of mutimessenger astronomy with the possible emergence of a new field, the astrophysics of neutrinos.

\section*{Acknowledgment}
The authors are thankful to D.V. Semikoz for useful suggestions.
M. Ribordy is supported by the Swiss National Research Foundation (grant PP002--114800).


\begin{thebibliography}{99}

\bibitem{ls5039} Aharonian F. et al., 2005, Science, 309, 746; 2006, A\&A, 460, 743.
\bibitem{lsi} Albert J. et al. 2006, Science, 312, 1771;  Acciari, V. A. et al., 2008, Ap.J., 679, 1427.
\bibitem{psrb} Aharonian A. et al., 2005, A\&A, 442, 1.
\bibitem{HESSJ0632} Hinton J.A. et al., 2008, arXiv:0809.0584.
\bibitem{cygx1} Albert J. et al., 2007, Ap.J., 665, L51.
\bibitem{mirabel} Mirabel I.F., 2006, Science, 312, 1759.
\bibitem{maraschi} Maraschi L., Treves A. 1981, MNRAS, 194, 1.
\bibitem{tavani97} Tavani M., Arons J., 1997, Ap.J., 477, 439.
\bibitem{bosch-ramon} Bosch-Ramon V. et al., 2006, A\&A, 447, 263.
\bibitem{dubus} Dubus G., 2006, A\&A, 451, 9;
\bibitem{bednarek} Bednarek W., 2006, MNRAS, 368, 579; 
\bibitem{khangulyan} Khangulyan D. et al. 2008, MNRAS, 383, 467.
\bibitem{romero} Romero G. et al. 2007, A\&A, 474, 15.
\bibitem{chernyakova06a} Chernyakova M., Neronov A. \& Walter R., 2006a, MNRAS, 372, 1585.
\bibitem{chernyakova06b} Chernyakova M., Neronov A., et al.  MNRAS, 2006b,  367, 1201;
\bibitem{lsi_burst} De Pasquale  M. et al., 2008, GCN 8209.
\bibitem{dhawan} Dhawan V., Mioduszewski A., Rupen M.   in Proc. of VI Microquasar workshop (Como).
\bibitem{ribo08} Rib\'o M. et al., 2008, A\&A, 481, 17. 
\bibitem{aharonian07n} Aharonian F.A., 2007, Talk at Neutrino 06, Santa Fe, New Mexico, June 13-19, 2006astro-ph/0702680.
\bibitem{aharonian06n} Aharonian F.A., Anchordoqui L.; Khangulyan D.; Montaruli T., 2006, J.Phys.: Conf. Series, 39, 408.
\bibitem{orellana} Orellana M.; Bordas P.; Bosch-Ramon V.; Romero G. E.; Paredes J. M., 2007, A\&A,476.
\bibitem{halzen} Torres D.F.; Halzen F., 2007, A.Ph., 27, 500.
\bibitem{kappes} Kappes A.; Hinton J.; Stegmann C.; Aharonian F. A., 2007, Ap.J., 656, 870.
\bibitem{christiansen} Christiansen H.R.; Orellana M.; Romero G.E., 2006, Phys.Rev. D, 73, 3012.
\bibitem{gallant94} Gallant Y.A., Arons J., 1994, Ap.J., 435, 230.
\bibitem{horns06} Horns D., Aharonian F., Santangelo A., Hoffmann A.I.D., Masterson C., 2006, A\&A, L51.
\bibitem{bednarek03} Bednarek W., 2003, A\&A, 407, 1.
\bibitem[Waters et al.(1988)]{waters88}
Waters L.~B.~F.~M., van den Heuvel E.~P.~J., Taylor A.~R., Habets 
G.~M.~H.~J., Persi P., 1988, A\&A, 198, 200
\bibitem[Mart{\'{\i}} \& Paredes(1995)]{mp95} 
Mart{\'{\i}} J., Paredes J.~M., 1995, A\&A, 298, 151 
\bibitem[Porter(1998)]{porter98} 
Porter J. M., 1998, A\&A, 333, L83
\bibitem[Blumenthal \& Gould(1970)]{bg70} 
Blumenthal G.~R., Gould R.~J., 1970, Rev.\ Mod.\ Phys., 42, 237 
\bibitem[Kelner, Aharonian \& Bugayov]{kelner} Kelner S. R.; Aharonian F. A.; Bugayov V. V.,  2006, Phys. Rev. D, 74, 4018.
\bibitem[Chernyakova, Neronov \& Walter(2006)] {chernyakova06}
Chernyakova M., Neronov A., Walter R., 2006, MNRAS, 372, 1585.
\bibitem{veritas} Acciari, V. A. et al.,2008, Ap.J., 679, 1427.
\bibitem{km3net}KM3NeT coll., see {\tt http://www.km3net.org/publications.php}
\bibitem{icecube} {\tt http://icecube.wisc.edu}
\bibitem[Achtenberg et al.(2007)]{amanda} Achenberg A., et al., 2007, Phys.Rev.D75:102001,2007. 
\bibitem{a2-7yrs}T.~I.~Collaboration, arXiv:astro-ph/0809.1646.
\bibitem{conv_tig} P.~Gondolo, G.~Ingelman, M.~Thunman, Astropart. Phys. {\bf 5}  309, 1996.
\bibitem{conv_lip} P.~Lipari, Astropart. Phys. {\bf 1} 195, 1993.
\bibitem{volkova}  L.V.~Volkova, Sov. J. Nucl. Phys. {\bf 31} 784, 1980.
\bibitem{antares-paper} G.~Anton, Nucl. Phys. B (Proc. Suppl.) {\bf 143} 351, 2005.
\bibitem{sendai}T.~Montaruli {\it et al.}in Proc. of 10th Int. Conf. on Topics in Astropart. and Underground Phys. (TAUP) 2007, Sendai, Japan, astro-ph/0712.3524.
\bibitem{prem}A.M. Dziewonski, D.L. Anderson, Phys. Earth Planet. Inter. 25, 297 (1981).
\bibitem{mrsg}A. D. Martin, W. J. Stirling, and R. G. Roberts,  Phys. Lett. B 354, 155 (1995); Int. J. Mod. Phys. A 10, 2885 (1995); Phys. Rev. D 51, 4756, (1995).
\bibitem{dima}IceCube coll., Juan-de-Dios Zornoza et al., in Proc. 30th ICRC, M\'erida (2007), arXiv:0711.0353.
\bibitem{composite-ribordy}M.~Ribordy, Nucl.\ Instrum.\ Meth.\  A {\bf 574}, 137 (2007).
\bibitem{liparistanev}P. Lipari and T. Stanev, Phys. Rev. D 44, 3543 (1991).
\bibitem{stanevgaisser}T.~K.~Gaisser and T.~Stanev, Phys.\ Rev.\  D {\bf 30}, 985 (1984).
\bibitem{halzen-newphys}M.~C.~Gonzalez-Garcia, F.~Halzen and M.~Maltoni, Phys.\ Rev.\  D {\bf 71} (2005) 093010.
\bibitem{gqrs}R.~Gandhi, C.~Quigg, M.~H.~Reno and I.~Sarcevic, Phys.\ Rev.\  D {\bf 58} (1998) 093009. Astropart.Phys.5:81-110,1996. 
\bibitem{halzen-milagro}Halzen, Kappes, Murchada
\bibitem{simul}J.~Braun, J.~Dumm, F.~De Palma, C.~Finley, A.~Karle and T.~Montaruli, Astropart.\ Phys.\  {\bf 29} (2008) 299.
\bibitem{mmc}D.~Chirkin and W.~Rhode, arXiv:hep-ph/0407075.
\end{thebibliography}
\end{document}